\documentclass[multphys,vecphys]{svmult}

\usepackage{makeidx}         % allows index generation
\usepackage{graphicx}        % standard LaTeX graphics tool
\usepackage{multicol}        % used for the two-column index
\usepackage[bottom]{footmisc}% places footnotes at page bottom
\makeindex             % used for the subject index
\begin{document}

\title*{Imprint of galaxy formation and evolution on globular cluster
properties}
\titlerunning{Imprint of galaxy formation on GCs}
%\author{Kenji Bekki\inst{1}}
\author{Kenji Bekki}
\institute{School of Physics, University of New South Wales, Sydney 2052, Australia
\texttt{bekki@bat.phys.unsw.edu.au}}
\maketitle

\begin{abstract}
We discuss the origin of physical properties of globular cluster
systems (GCSs) in galaxies in terms of galaxy formation and evolution
processes. Based on numerical simulations of dynamical evolution
of GCSs in galaxies, we particularly discuss (1) the origin of
radial density profiles of GCSs, (2) kinematics of GCSs in elliptical
galaxies, (3) transformation from nucleated
dwarf galaxies into GCs (e.g., omega Centauri),
and (4) the origin of GCSs in the Large Magellanic 
Cloud (LMC).
\end{abstract}

\section{Numerical archeology}
Based on penetrative analysis of 
metal-poor  halo stars and globular
clusters (GCs) in the Galaxy, 
two canonical Galaxy formation scenarios 
-- the monolithic collapse scenario \cite{ELS62}
%(Eggen, Lynden-Bell, \& Sandage 1962)
and 
the accretion/merging one \cite{SZ78}-- 
%(Searle \& Zinn 1978)-
were proposed,
that have long been influential for later observational
and theoretical studies of disk and elliptical galaxies.
Although observational studies of stellar halos 
in galaxies 
beyond the Local Group of galaxies
have just recently started revealing structural and chemical 
properties of the halos 
%(e.g., Dalcanton \& Bernstein 2002;
%Zibetti et al. 2004), 
\cite{DB02}\cite{Z04},
physical properties of globular cluster systems (GCSs) in
these galaxies have long been investigated 
in much more details 
%(e.g., Harris 1991).
\cite{H91}.
Wide-field imaging and spectroscopic studies
with large ground-based  telescopes (e.g., Keck 10m)
have recently revealed GCS structures and kinematics in 
galaxies with different Hubble types \cite{BS06}. 
%(Brodie \& Strader 2006
%for a recent review).
Furthermore a growing number of
theoretical/numerical studies have recently been accumulated 
which have investigated 
dynamical and chemical properties of 
GCs and GCSs   based on  admittedly realistic
and self-consistent models of GC formation 
during galaxy formation and evolution \cite{B02}\cite{GN06}.
%(e.g., Bekki et al. 2002; Gnedin 2006). 
In this review paper, we therefore 
try to derive physical meanings from 
the selected four observed  properties of GCs and
GCSs
by comparing our numerical simulations of GC and GCS formation
with  latest observations.

\begin{figure}
\centering
\includegraphics[height=10cm]{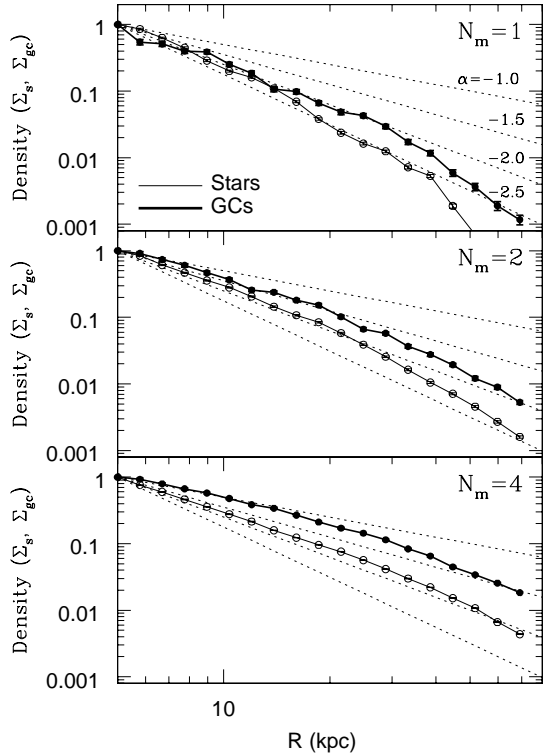}
\caption{
Dependences of projected number distributions of stars (thin)
and GCs (thick) in merger remnants (i.e., elliptical galaxies)
on  the total number of major 
merger events ($N_{\rm m}$) which an elliptical 
experienced during its formation (BF06).
For clarity, the density distributions are normalized to
their central values.
Thin dotted lines represent power-law slopes ($\alpha$)
of $\alpha$ = $-2.5$, $-2.0$, $-1.5$, and $-1.0$. 
Note that the density profiles of GCSs become flatter
for larger $N_{\rm m}$, i.e. more mergers.
}
\label{fig:1}       % Give a unique label
\end{figure}

\section{GCS density profiles}
It has long been known that the radial density profiles of
GCSs in elliptical galaxies (Es)
vary with the total luminosities of their host galaxies \cite{AZ98}.
%(e.g., Ashman \& Zepf 1998).
If the projected GCS density profiles 
in Es with $V$-band absolute magnitudes of $M_{\rm V}$ 
are fitted to the power-law ones like 
${\Sigma}_{\rm gc} \propto R^{{\alpha}_{\rm gc}}$,
where $R$ is the distance from the center of the host galaxy of a GCS,
the power-law index ${\alpha}_{\rm gc}$ is smaller (i.e.,
the profiles are steeper) for larger $M_{\rm V}$ (i.e., fainter Es).
Although two physical mechanisms -- GC destruction by galactic tidal
fields \cite{B98}\cite{V03} 
%(e.g., Baumgardt 1998; Vesperini et al. 2003)
and dynamics of galaxy merging 
(Bekki \& Forbes 2006; BF06; \cite{BF06}) -- 
have been so far proposed for the origin of GCS density profiles,
we focus on the latter case in this paper.

BF06
numerically investigated the structural properties
of GCSs in Es formed from a sequence of
major dissipationless galaxy merging  
and thereby found  that the radial density profiles of GCSs
in Es  become progressively flatter as the galaxies
experience more major merger events (See Figure 1).
The simulated  profiles of GCSs  
are found to be well described as power-laws
with ${\alpha}_{\rm gc}$ ranging from $-2.0$ to $-1.0$
in Es.
They are flatter than,
and linearly proportional to, the slopes (${\alpha}_{\rm s}$)
of the stellar density profiles.
By applying a reasonable scaling relation between luminosities and sizes of galaxies
to the simulation results,
BF06 showed that ${\alpha}_{\rm gc} \approx -0.36 M_{\rm V}-9.2$,
$r_{\rm c} \approx -1.85 M_{\rm V}$,
and ${\alpha}_{\rm gc} \approx 0.93 {\alpha}_{\rm s}$.
These correlations between GCS profiles and their host galaxy luminosities
are consistent reasonably well observations 
\cite{H86}\cite{AZ98}
%(e.g., Harris 1986; Ashman \& Zepf 1998),
which suggests that  the origin of structural non-homology of GCSs
in Es can be understood in terms of the  growth
of Es via major dissipationless galaxy merging.

\begin{figure}
\centering
\includegraphics[height=5cm]{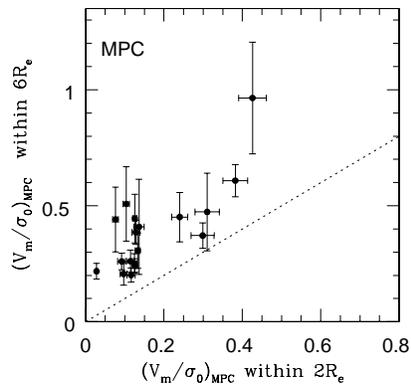}
\caption{
Correlations between ${(V_{\rm m}/{\sigma}_{0})}_{\rm MPC}$
estimated for $R$ $\le$ $2R_{\rm e}$ and 
those for $R$ $\le$ $6R_{\rm e}$  for metal-poor clusters (MPCs).
These correlations are derived from 18 results of 6 major merger
models with three different projections \cite{B05}.
}
\label{fig:2}       % Give a unique label
\end{figure}

\section{GCS kinematics in E/S0s}
Recent observations on GCS kinematics in E/S0s have
revealed that GCS kinematics can be quite diverse:
GCSs in some galaxies like M87 
%(e.g., Kissler-Patig \& Gebhardt 1998),
\cite{KG98}
NGC 4472 
\cite{Z00}
%(e.g., Zepf et al. 2000), 
and NGC 5128 
\cite{P04}
%(Peng et al. 2004)
show rotation whereas those in some galaxies
such as NGC 1399 
\cite{R04}
%(e.g., Richtler et al. 2004)
do not.
Bekki et al. (2005; B05; \cite{B05}) first tried to understand
the observed diversity in GCS kinematics 
by numerically investigating GCS kinematics of E/S0s formed from
major/minor galaxy merging.
B05 demonstrated that 
both metal-poor cluster (MPCs) and metal-rich ones (MRCs) in
Es formed from major mergers can exhibit
significant rotation at large radii ($\sim$20 kpc) due to
the conversion of initial orbital angular momentum into
intrinsic angular momentum of the remnant. 

Based on a wide parameter study of galaxy mergers,
B05 found that MPCs show higher
central velocity dispersions than MRCs for most major
merger models. $V_{\rm m}/{\sigma}_{0}$ (where $V_{\rm m}$
and ${\sigma}_{0}$, are the GCS maximum rotational velocity and
central velocity dispersion, respectively)
ranges from 0.2--
1.0 and 0.1--0.9 for the MPCs and  MRCs respectively, within
$6R_{\rm e}$ for the remnant elliptical. 
Figure 2 shows an interesting result that 
does not depend on merger parameters:
$V_{\rm m}/{\sigma}_{0}$ of MPCs 
within $6R_{\rm e}$ are  greater than those of MPCs 
within $2R_{\rm e}$.

B05 also revealed the alignment of the major axes
in 2D distributions between stars, GCs, and dark matter halos
in the simulated Es. 
The aligned major axis between stars, GCs, and dark matter  appears to be
one of the principal characteristics of Es 
formed by major merging, which
implies that {\it observational studies on 2D distributions of GCSs
in Es can tell us about the shapes of their host dark matter halos}.
We also showed in this meeting that the total masses of E/S0s estimated
from the GCS kinematics can be much closer to the real masses
than those  from the PNe systems owing to less anisotropic 
velocity dispersion in GCSs, in particular, for face-on S0s.
This suggests that (1) GCSs are better mass-estimators in 
E/S0s and (2) kinematical data sets of PNe systems 
in E/S0s \cite{R03} should be more
carefully interpreted for the total masses of E/S0s.
Although these simulations are not based on cosmological N-body ones,
these results may well provide new insight on the origin of
the observed diversity in GCS kinematics of E/S0s.

\begin{figure}
\centering
\includegraphics[height=5cm]{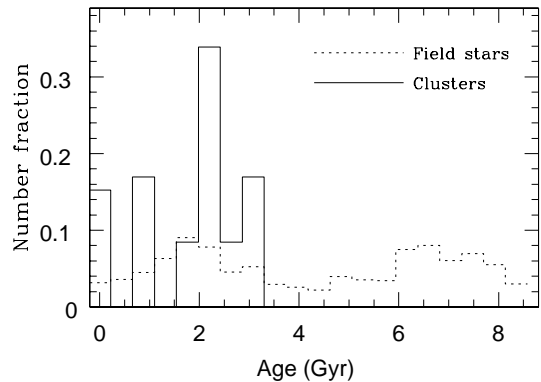}
\caption{
The simulated age distributions of field stars (dotted) and clusters
(solid) in the LMC at  the present epoch \cite{BC05}.
For convenience, the normalized fraction of stars
in each age bin is shown.
}
\label{fig:3}       % Give a unique label
\end{figure}

\section{The age gap problem in the LMC}

Possible candidates of young and metal-rich GCs were discovered in
interacting and merging galaxies \cite{WS95},
%(e.g., Whitmore \& Schweizer 1995),
and physical properties of these GCs have been discussed in
different contexts of galaxy and GC formation,
such as the observed  color bimodality in GCSs in Es \cite{AZ98},
%(Ashman \& Zepf 1998), 
the birth rate of young GCs as a function of time
in M82 
%(e.g., de Grijs et al. 2003),
\cite{d03}
and the age-metallicity relation of GCs in nearby interacting galaxies like
the LMC and the SMC 
(e.g., Bekki et al. 2004; B04; \cite{B04}).
In this paper, we focus on the LMC's GCS in which 
nearly all GCs are either very old ($\sim 13$ Gyr) 
or younger than $3-4$ Gyr --  the ``age-gap'' 
problem 
%(e.g., Da Costa 1991; Geisler 2006).
\cite{D91}\cite{G06}.

B04 and Bekki \& Chiba (2005; \cite{BC05}) 
challenged this age gap problem by investigating
chemodynamical evolution of the LMC interacting both with the Galaxy
and the SMC for a long time scale ($\sim 9$ Gyr).
They found that
the first close encounter
between the LMC and the SMC about 4 Gyr ago
was the beginning of a period of strong tidal interaction witch
likely induced dramatic gas cloud collisions, leading to an enhancement
of the GC formation  which has been sustained by
strong tidal interactions to the present day.
Figure 3, showing the simulated age distributions of field stars
and GCs in a model, reveals that GC formation can be reactivated
about $3-4$ Gyr ago, when the LMC can start its strong tidal interaction
with the SMC. 
These results imply that the origin of the age gap can be closely
associated with interaction histories  of the LMC, the SMC, and the Galaxy.

\begin{figure}
\centering
\includegraphics[height=10cm]{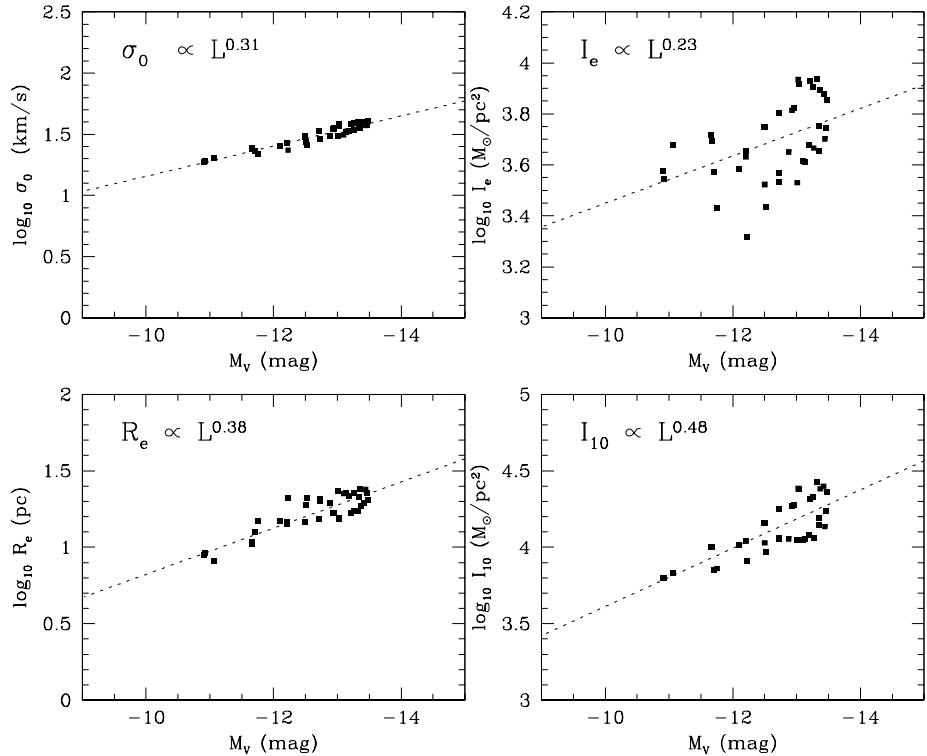}
\caption{
Correlations of structural and kinematical parameters with $M_{\rm V}$ ($V-$band 
absolute magnitude) for the VMSCs in 40 models
\cite{B04}.
Projected central velocity dispersion
($\sigma_{0}$; {\it upper left}), half-light-averaged surface brightness
($I_{e}$; {\it upper right}), effective radius ($R_{e}$; {\it lower left}),
and central surface brightness ($I_{10}$; {\it lower right}) are plotted against 
$M_{\rm V}$. Here the central surface brightness $I_{10}$ is expressed as
$0.1L/\pi/{R_{10}}^{2}$, where $L$, $R_{10}$ are the total luminosity of a VMSC
and the radius within which 10\% of $L$ is included, respectively. The best fit
scaling relation for the VMSCs is derived for  each panel using the least square 
fitting method and described as a {\it dotted} line with the derived relation
(e.g., ${\sigma}_{0}$ $\propto$ $L^{0.31}$).
}
\label{fig:4}       % Give a unique label
\end{figure}

\section{Very massive star clusters}

Very massive star clusters  (VMSCs) such as $\omega$ Centauri,
ultra-compact dwarfs (UCDs), and massive nuclear star clusters
have unique characteristics that are quite different from
those of ``normal'' GCs. 
For example,  UCDs discovered in
the Fornax and the Virgo clusters of galaxies
%(Drinkwater et al. 2003; Jones et al. 2006)
\cite{D03}\cite{J06}\cite{M04}
have intrinsic sizes of less than  100\,pc, 
and have absolute
$B-$band magnitudes ranging from $-13$ to $-11$\,mag,
which is more than 2 magnitudes brighter than the most massive GC
in the Galaxy (i.e., $\omega$ Cen).
Two physical mechanisms for the VMSC formation have been so
far proposed: The ``Galaxy threshing'' scenario 
%(Bekki et al. 2001; 2003) 
\cite{B01}\cite{B03}
in which
VMSCs originate from nuclei of nucleated dwarf galaxies
and the ``cluster merging'' one in which VMSCs are formed
from merging of smaller star clusters in tidal tails of merging galaxies
%(Fellhauer \&  Kroupa 2002).
\cite{FK02}.
Although  more details on physical properties of VMSCs have been
recently revealed
%(e.g., Hilker et al. 2004; Ha\c segan et al. 2005),
\cite{H04}\cite{H05},
it remains unclear which of the two scenarios is more convincing
for the origin of VMSCs.

Here we discuss what we can learn from VMSC properties {\it if they
were previously nuclei of nucleated galaxies.} Both dissipationless
%(e.g., Tremaine et al. 1975;  Capuzzo-Dolcetta \& Tesseri 1999)
\cite{T75}\cite{CT99}
and dissipative
%dissipative (Bekki et al. 2006a)
\cite{B06a}
formation scenario of stellar galactic nuclei
have provided some interesting predictions on nuclear properties
of galaxies 
%(e.g., Bekki et al. 2004).
\cite{BCDS04}.
Figure 4 shows the scaling relations between different physical
parameters of merger remnants of star clusters initially with
the observed GC scaling relations 
\cite{D97}.
%(Djorgovski et al. 1997).
The fact that the simulated relations deviate from
the GC's ones implies that {\it the scaling relations of VMSCs
can be used for understanding whether the stellar nuclei
can be formed from merging of many smaller clusters in
the central regions of galaxies}
\cite{C06}.
%(e.g., C\^ote et al. 2006). 
The dissipative nucleus formation model 
%(Bekki et al. 2006a)
\cite{BCDS04}
has predicted the spread of ages and metallicities 
in stellar populations of stellar galactic nuclei
and accordingly  can be discussed in the context of the observed
spread of ages and metallicities in $\omega$ Cen 
%(e.g., Hilker et al. 2004).
\cite{H04}.
Thus dynamical and chemical properties of VMSCs can
tell us about nucleus formation histories in galaxies.

\section{Future works: Hierarchical galaxy formation and GCSs}

Thus, structural, kinematical, and chemical properties of GCSs in
galaxies have fossil  information on dynamics of major/minor galaxy merging
(e.g, angular momentum redistribution processes in merging),
interaction histories of galaxies,  and the  formation histories of stellar
galactic nuclei in dwarfs.
Previous theoretical/numerical 
studies however did not  discuss so extensively
the observed correlations between GCS properties and their host ones
%(e.g., Brodie \& Strader 2006 for a recent review)
\cite{BS06}
in the context of a hierarchical clustering scenario of galaxy formation.
Several authors just recently have 
started their investigation on the GCS-host relations
based on semi-analytic models 
%(Beasley et al. 2002)
\cite{BB02}
and high-resolution numerical simulations in $\Lambda$CDM models
%(Rhode et al 2005; Bekki et al. 2006b). 
\cite{R05}\cite{B06b}.
A number of observed GCS-host relations, such as
the positive correlation between GCS metallicities and their
host galaxy luminosities 
%(Brodie \& Strader 2006),
\cite{BS06}
have not been clearly explained by any galaxy formation scenarios.
Since these GCS-host relations may have profound physical meanings
on galaxy formation and evolution,
it is doubtlessly worthwhile for future numerical simulations of
GCS formation to explore the origin of these relations.

%\label{sec:1}

%\begin{equation}
%\vec{a}\times\vec{b}=\vec{c}
%\end{equation}

%
% For built-in environments use
%
% BibTeX users please use
% \bibliographystyle{}
% \bibliography{}

\begin{thebibliography}{99.}
\bibitem{ELS62}
O.J. Eggen, D. Lynden-Bell,  A.R. Sandage:  ApJ, 
\textbf{136},
748 (1962)
\bibitem{SZ78}
L. Searle, R. Zinn:  ApJ, 
\textbf{225}, 357 (1978)
\bibitem{DB02}
J. Dalcanton, R.A. Bernstein: AJ, 
\textbf{123}, 1328 (2002)
\bibitem{Z04}
S. Zibetti, S.D.M. White, J. Brinkmann: MNRAS, 
\textbf{347}, 556 (2004)
\bibitem{H91}
W.E. Harris: ARA\&A, 
\textbf{29}, 543 (1991)
\bibitem{BS06}
J.P.  Brodie, J. Strader: ARA\&A in press (astro-ph/0602601) (2006)
\bibitem{B02}
K. Bekki, D.A. Forbes, M.A. Beasley,
W.J. Couch: MNRAS, 
\textbf{335}, 1176 (2002)
\bibitem{GN06}
O.Y. Gnedin:  in this volume  (2006)
\bibitem{AZ98}
K.M. Ashman, S.E. Zepf: in Globular cluster systems,
Cambridge, U. K. ; New York : Cambridge University Press (1998)
\bibitem{B98}
H. Baumgardt:  A\&A, 
\textbf{330}, 480 (1998)
\bibitem{V03}
E. Vesperini, S. E. Zepf, A. Kundu,   K.M. Ashman: ApJ, 
\textbf{593}, 760 (2003)
\bibitem{BF06}
K. Bekki, D.A. Forbes: A\&A, 
\textbf{445}, 485 (2006)
\bibitem{H86}
W.E. Harris: AJ, 
\textbf{91}, 822 (1986)
\bibitem{KG98}
M. Kissler-Patig, K. Gebhardt:  AJ, 
\textbf{116}, 2237 (1998)
\bibitem{Z00}
S.E. Zepf, M.A. Beasley et al:
AJ, \textbf{120}, 2928 (2000)
\bibitem{P04}
E.W. Peng, H.C. Ford, K.C. Freeman: ApJ, 
\textbf{602}, 705 (2004)
\bibitem{R04}
T. Richtler et al: AJ, 
\textbf{127}, 2094 (2004)
\bibitem{B05}
K. Bekki, M.A. Beasley, J.P. Brodie, D.A. Forbes,
MNRAS, 
\textbf{363}, 1211 (2005)
\bibitem{R03}
A.J. Romanowsky, N.G. Douglas et al:
Science, 
\textbf{301}, 1696 (2003)
\bibitem{WS95}
B.C. Whitmore, F. Schweizer: AJ, 
\textbf{109}, 960 (1995)
\bibitem{d03}
R. de Grijs, N. Bastian, H.J.G.L.  Lamers: MNRAS, 
\textbf{340}, 197 (2003)
\bibitem{B04}
K. Bekki, W.J. Couch, W. J. et al:
\textbf{610}, L93 (2004)
\bibitem{D91}
G.S. Da Costa: 
in Haynes R., Milne D., eds, Proc. IAU Symp. 148,
The Magellanic Clouds, Kluwer, Dordrecht, 
p183 (1991)
\bibitem{G06}
D. Geisler: in this volume (2006)
\bibitem{BC05}
K. Bekki, M. Chiba: MNRAS, 
\textbf{356}, 680 (2005)
\bibitem{D03}
M.J. Drinkwater, M.D. Gregg et al:
Nature, 
\textbf{423}, 519 (2003)
\bibitem{J06}
J.B. Jones, M.J. Drinkwater et al:
AJ, 
\textbf{131}, 312 (2006)
\bibitem{M04}
S. Mieske, M. Hilker, L. Infante:, A\&A, 
\textbf{418}, 445 (2004)
\bibitem{B01}
K. Bekki, W.J. Couch, M.J. Drinkwater:
ApJ, 
\textbf{552}, L105 (2001)
\bibitem{B03}
K. Bekki, W.J. Couch, M.J. Drinkwater, Y. Shioya:
MNRAS, 
\textbf{344}, 399 (2003)
\bibitem{FK02}
M. Fellhauer, P. Kroupa: MNRAS, 
\textbf{330}, 642 (2002)
\bibitem{H04}
M. Hilker, A. Kayser, T. Richtler, P. Willemsen:
A\&A, 
\textbf{422}, L9 (2004)
\bibitem{H05}
M. Ha\c segan et al: ApJ, 
\textbf{627}, 203 (2005)
\bibitem{T75}
S.D. Tremaine, J.P. Ostriker, L. Spitzer, Jr:
ApJ, 
\textbf{196}, 407 (1975)
\bibitem{CT99}
R. Capuzzo-Dolcetta, A. Tesseri: MNRAS, 
\textbf{308}, 961 (1999)
\bibitem{B06a}
K. Bekki, W.J. Couch, Y. Shioya:  ApJL, in press (astro-ph/064340) (2006)
\bibitem{BCDS04}
K. Bekki, W.J. Couch, M.J. Drinkwater, Y. Shioya, 
ApJ, 
\textbf{610}, L13 (2004)
\bibitem{D97}
S.G. Djorgovski, R.R. Gal et al:
ApJ, 
\textbf{474}, L19 (1997)
\bibitem{C06}
P. C\^ote  et al: 2006, ApJS in press (astro-ph/0603252) (2006)
\bibitem{BB02}
M.A. Beasley, C.M. Baugh et al:
MNRAS, \textbf{333}, 382 (2002)
\bibitem{R05}
K.L. Rhode, S.E. Zepf, M.R. Santos: ApJ, 
\textbf{630}, L21 (2005)
\bibitem{B06b}
K. Bekki,  H. Yahagi, D.A. Forbes: submitted to ApJL  (2006)


%%%%%%%%%%



\end{thebibliography}
%
% Non-BibTeX users please follow the syntax
% the syntax of "referenc.tex" for your own citations
%\input{referenc}
%%%%%%%%%%%%%%%%%%%%%%%%%%%%%%%%%%%%%%%%%%%%%%%%%%%%%%%%%%%%%%%%%%%%%%  }

%%%%%%%%%%%%%%%%%%%%%%%%%%%%%%%%%%%%%%%%%%%%%%%%%%%%%%%%%%%%%%%%%%%%%%

\printindex
\end{document}